\documentstyle[twocolumn]{article}

\begin{document}
\title{Ram-Ppressure Effects on Dense Molecular Arms
in the Central Regions of Spiral Galaxies by Intracluster Medium}
\author{Makoto H{\sc IDAKA}, and Yoshiaki S{\sc OFUE}\\
{\it Institute of Astronomy, The University of Tokyo,
    Mitaka, Tokyo, 181-0015} \\
	 sofue@ioa.s.u-tokyo.ac.jp}
\maketitle

\def\ficm{f_{\rm ICM}}
\def\fism{f_{\rm ISM}}
\def\rhoicm{\rho_{\rm ICM}}
\def\rhoism{\rho_{\rm ISM}}
\def\nicm{n_{\rm ICM}}
\def\nism{n_{\rm ISM}}
\def\cosa{{\rm cos}\alpha}
\def\sina{{\rm sin}\alpha}
\def\vrot{v_{\rm rot}}
\def\dv{\delta v}
\def\kms{ km s$^{-1}$ }
\def\be{\begin{equation}}
\def\ee{\end{equation}}
\def\HI{H {\sc i}}

\begin{abstract}
We investigated ram-pressure effects by an intracluster wind on an inner disk
of spiral galaxies by a hydrodynamical simulation.
Even if the wind is mild and not strong enough to strip the gas disk, the ram
pressure disturbs orbits of the inter-arm gas significantly.
This results in asymmetric dense molecular arms in the inner few kpc region
of a galaxy.
This mechanism would explain the asymmetric CO gas distributions in the
central regions often observed in Virgo spirals.

Key words: galaxies:   spiral --- galaxies: kinematics and dynamics
--- galaxies: ISM --- galaxies: cluster of --- intergalactic matter
\end{abstract}

\section{Introduction}

Ram-pressure by intracluster medium (ICM) causes the stripping of interstellar
matter (ISM) from galaxies (Farouki,  Shapiro 1980; Kritsuk 1984;
Gaetz et al. 1987; Balsara et al. 1994; Sofue 1994; Quilis et al. 2000).
It also produces a disturbed distribution of ISM in galaxies, such as
head-tail H {\sc i} structures as observed in Virgo galaxies
(Cayatte et al.  1990; Vollmer et al. 2000, 2001; Phookun et al. 1993, 1995).
The ram-pressure effect has, thus, been discussed mainly in relation to \HI\ gas
stripping and outer disk structures.
However, little attention has been paid to its effect on the inner molecular
disk and arms.
Only a few authors have discussed the ram effect on the spiral structure
(Tosa 1994) and molecular clouds (Kenney et al. 1990; Sofue 1994).

Kenney et al. (1990) have shown that Virgo galaxies often exhibit asymmetric
inner molecular disks.
A recent high-resolution CO-line survey of Virgo cluster galaxies such as
NGC 4254 and NGC 4654 (Sofue et al. in preparation) has revealed  that
some of them show a significant asymmetry of the inner molecular disks and arms.
It is not clear if such an inner deformation of dense gas disks can be
produced by ram pressure effect, which is thought to be the cause of
their deformed \HI\ envelops.
Since these two galaxies have no massive companion that can disturb such
inner disks, and they both show a head-tail \HI\ outer structure (Phookun et al.
1993, 1995), the inner deformation of molecular disks could also be due to
the ram-pressure effect, while such an inner ram effect has not yetbeen 
investigated.

If the wind is very strong, such as assumed by Quilis et al. (2000)
for the core of a rich cluster with an ICM density of $\sim 3 \times 10^{-3}$
atoms cm$^{-3}$ and  a wind velocity higher than $\sim2000$ \kms, 
the ISM of any galaxies would be completely stripped.
On the other hand, if the wind is mild, in such a case as for
the Virgo cluster, where the ICM density is of the order of
$10^{-3} - 10^{-4}$ and the velocity is $\sim 1000$ \kms, outer \HI\ envelopes
are deformed to produce head-tail structures (Vollmer et al. 2000, 2001).
In current simulations, such as those by Vollmer et al., which were aimed 
at gas stripping and tailing of the outer \HI\ disks, the detailed behavior
of the inner disk gas inside $\sim 10$ kpc was not well understood because of
the resolution.

In the present paper, we consider a mild ICM wind, and discuss its effect
on the inner disk gas based on 2D hydrodynamical simulations with higher 
resolution than those aimed at outer \HI\ stripping, as above.
If we simply apply the ram-stripping condition to an azimuthally
structure-less gas disk, the ram pressure would hardly affect the inner disk.
However, if we consider a spiral structure with an arm-to-interarm density
contrast, it may happen that the ram-pressure can affect the low-density
interarm gas.
We consider here the possibility that the ram-pressure can affect
the dense molecular gas within the central  few kpc region of spiral
galaxies through disturbances of the orbits of inter-arm gas,
even if the ICM wind is not strong enough to strip the disk gas.

\section{Ram-Pressure Force on the Arm and Interarm Gases}

The component of the ram force parallel to the galactic plane exerted by an
intergalactic wind on a gas element is given by
\be
\ficm \sim \rhoicm ~s~ \dv^2  ~\cosa ~\sina,
\ee
where $\rhoicm$ and $\rhoism$ are the gas densities of ICM and ISM,
respectively, $s$ is the surface area of the element, and
$\alpha$ is the angle between the wind direction and the galactic plane
(Farouki,  Shapiro 1980; Kritsuk 1984; Sofue 1994; Tosa 1994).
The motion of the undisturbed ISM element is governed by the gravitational
force, which is approximately equal to the centrifugal force,
\be
\fism \sim \rhoism ~d ~s ~ \vrot^2/R,
\ee
where $d$ is the thickness of the gas disk, and $R$ is the galactocentric
radius of the element.
Now, the ratio of $\ficm$ to $\fism$ is given by
\be
\eta \sim {\nicm \over \nism}
	{R \over d} \bigl({\dv \over \vrot}\bigr)^2~ \cosa ~\sina,
\ee
where $\nicm$ and $\nism$ are the number densities of hydrogen atoms
of the ICM and ISM, respectively.
If $\eta$ exceeds unity, the ram force can disturb the ISM motion, while
if it is smaller than unity, the ISM motion is little affected.

Let us consider the inner part of a galaxy at
$R \sim 5 $ kpc with $d\sim 100$ pc,
rotating at  $\vrot \sim 200$ \kms.
For an ICM wind with  $\nicm \sim 10^{-4}$ cm$^{-3}$,
$V_{\rm ICM} \sim 10^3$ \kms, and $\alpha \sim 45^\circ$,
we obtain the ratio to be
\be
\eta \sim 0.6 ~\nism^{-1} ~[{\rm H~cm^{-3}}].
\ee
This relation implies that the ISM is stripped if $\nism\ll 1$ H cm$^{-3}$,
since the force perpendicular to the galactic plane is of the same order.
We stress, however, that the relation indicates that the orbits of
the gas within the disk plane is significantly disturbed if $\nism \sim 1$
H cm$^{-3}$, even if the wind is not strong enough to strip the gas.
This may indeed apply to the inter-arm ISM in the inner disk within a few kpc
radius.
On the other hand, high-density galactic shocked arms, where
$\nism \gg 1$ H cm$^{-3}$, would be hardly disturbed.
We, thus, anticipate that, even if the ICM wind is mild, not strong enough
to strip the disk, the interarm gas in the inner disk would be significantly
disturbed, which may result in  deformed galactic shock waves.
Since $\delta v$ is greater on the head-wind side of the rotation axis compared
to the following-wind side, the deformed shock waves could be asymmetric
with respect to the rotation axis.

Figure 1 illustrates the ram-deformation mechanism of the
dense molecular arms, and how a lopsided spiral pattern is created.
Since the density of the inter-arm gas (A and A$'$ in Fig. 2) is 
much lower than 
the average density of the disk, the ram force by the ICM wind (thin lines) 
easily disturbs the orbits of inter-arm gas (thick curved arrows).
The gas on the distorted orbits encounters density waves (dashed spirals)
at different places (B and B$'$) from those expected for undisturbed orbits
(dashed lines), and produces deformed dense molecular arms (thick spirals).
In the next sections we discuss a numerical simulation of the ram deformation
of spiral arms in order to understand whether this mechanism can 
indeed create deformed shocked arms, and how the deformed arms 
look like in realistic model disks.

--- Fig. 1 ---

\section{Numerical Simulation}

\subsection{Basic assumptions}
The ram-pressure acceleration per unit mass is given by
\begin{equation}
\vec{a}_{\rm{ram}} = C n_{\rm{ICM}} \left| \vec{V}_{\rm{ICM}}
- \vec{v}_{\rm{rot}} \right| \left( \vec{V}_{\rm{ICM}} - \vec{v}_{\rm{rot}}
\right),
\end{equation}
where $n_{\rm{ICM}}$ is the ICM density, $\vec{V}_{\rm{ICM}}$ is the ICM
velocity with respect to the galaxy, $\vec{v}_{\rm{rot}}$ is the circular
velocity of the clouds, and $C$ is evaluated to be on 
the order of $\Sigma^{-1}$ (e.g., Tosa 1994).

We assume that the galactic disk is thin and faces the ICM wind everywhere,
being not shielded by neighboring clouds.
Because this assumption would not apply to an edge-on wind, we consider here
a wind with  a mild inclination.
We consider the inner disk of a galaxy, where the gravitational potential
is deep and the ISM is dense enough so that stripping does not
occur, as discussed in the previous section, and we treat a 2D disk in a fixed
rotating potential.

For the velocity of a galaxy in the ICM $\vec{V}_{\rm{ICM}}$
and the ICM density $n_{\rm{ICM}}$,
we adopt three values for each parameter.
The galaxy's velocity, $\vec{V}_{\rm{ICM}}$, is taken to be
530, 1000, and 1500 \kms,
where the first value is suggested by Phookun,  Mundy (1995) for NGC 4654
in the  Virgo cluster.
The ICM density, $n_{\rm{ICM}}$ is taken to be
$1 \times 10^{4}$ cm$^{-3}$,
$ 5 \times 10^{-4}$ cm$^{-3}$, and $1 \times 10^{-3}$ cm$^{-3}$, where
the first value is typical for the intergalactic density.

\subsection{Numerical Method}

For simplicity, we assumed that the interstellar gas is ideal, inviscid,
and compressible. We used a freely downloadable and usable
hydrodynamical code, VH-1 (Blondin, Lufkin 1993).
This is a multidimensional hydrodynamics code for an ideal compressible fluid
written in FORTRAN, developed by the numerical astrophysics group at the
University of Virginia based on  the Piecewise Parabolic Method
Lagrangian Remap (PPMLR) scheme of Colella, Woodward (1984).
The PPMLR has the advantage of
maintaining contact discontinuities without the aid of a contact steepener,
and is sufficiently good to be applied to a galactic-scale 
hydrodynamical simulation.
The code does not take into account the gas's self-gravity,
artificial viscosity, variable gamma equation of state, and radiative heating
and/or cooling. 
The self-gravity of the gas is not taken into account, because we consider a
case where the gas mass is not so much as to contribute to the density wave 
potential by the stellar disk, and also because we do not intend to discuss 
such processes as clumping of gas and cloud formation in the arm and 
inter-arm regions.
The interstellar gas can be assumed to be isothermal, as is assumed here,
while if cooling is taken into account, the shocked gas arms would become
much denser than those calculated below.
However, all such neglection would not affect the physical essence of the
present study, which was aimed at simulating ram-deformation of the orbits of 
inter-arm gas and resulting disturbed shocked arms, as illustrated in figure 1.

\subsection{Gravitational Potential}
We implicitly give a gravitational potential, which comprises the following
two terms:
(i) a static axisymmetric potential, and
(ii) a nonaxisymmetric, rotating bar potential.
The potential is expressed by
\begin{equation}
\Phi (R, \phi ) = \Phi_0 (R) + \Phi_1 (R, \phi ).
\end{equation}
We adopt a ``Toomre disk'' (Toomre 1981) potential for the axisymmetric
component as given by 
\begin{equation}
\Phi_0 (R) = - \frac{c^2}{a} \frac{1}{(R^2 + a^2)^{1/2}}, \label{eq:4.5}
\end{equation}
where $ a $ is the core radius and $c = v_{\rm{ max}} (27/4){}^{1/4} a$.
Through our numerical simulation, we fixd the core radius
and maximum circular velocity to be $a = \sqrt{2}$ kpc and $v_{\rm {max}}
= 200$ \kms.

The nonaxisymmetric potential was taken from Sanders (1977), assuming rigid
rotation at a pattern speed, $\Omega_{\rm p}$, which has the form
\begin{equation}
\Phi_1 (R, \phi ) = \varepsilon \frac{aR^2}{(R^2 + a^2)^{3/2}} \Phi_0(R)
\cos 2\left(  \phi - \Omega_p t \right), \label{eq:4.6}
\end{equation}
where $\varepsilon$ is the strength of the bar of the order of
$\varepsilon = 0.15$.
Spiral shocked arms of gas are produced by this potential.

\subsection{Initial Conditions}
Initially, we set $256 \times 256$ two-dimensional cells corresponding to
12.8 kpc $\times 12.8$ kpc field, while setting the field center at
the coordinates origin.
The initial number density was taken to be 5 cm${}^{-1}$ the inner disk at
$R \ge 8$ kpc
disk, and 1 cm${}^{-1}$ at $R> 8$ kpc.
The initial rotation velocity of each gas cell was set so that the centrifugal
force would balance the gravitation.
The bar pattern speed,  $\Omega_{\rm p}$, was taken to be 23 \kms, and
the strength of the bar $\varepsilon$ was taken to be 0.10.

\section{Ram-Pressure Deformation of Dense Molecular Arms}

\subsection{Deformation of Inner Spiral Structure}
Figure 1 shows the result of a simulation including ram-pressure effects for the
various parameter combinations, as described in the previous section.
Both the ISM and spiral pattern rotate counterclockwise, and the ICM wind 
blows from left to right.
The simulation shows that the orbits of the diffuse inter-arm gas
are easily disturbed by the ram force, which results
in a significant displacement of galactic shock waves from their undisturbed
symmetric positions, as illustrated in figure 1.

Highly asymmetric dense spiral arms in the central region are produced 
by this mechanism if the wind speed is higher than
$\sim 1000$ \kms\ and the ICM density is greater than
several $10^{-4}$ H cm$^{-3}$.
A head-tail structure of dense gases slanted to the ICM wind,
like NGC 4654 nucleus, can be produced by this mechanism,
if $n_{\rm{ICM}} \times V_{\rm{ICM}}^2$ is
greater than $\sim 3 \times 10^{12}$ cm${}^{-1}$ s${}^{-2}$.
One spiral arm on the downstream side is prominent, reproducing the lopsided
arms, as observed in NGC 4254 and NGC 4654.

--- Fig. 2 --- 

\subsection{Deformed Molecular Arms}

\def\fmol{f_{\rm mol}}
\def\rhohtwo{\rho_{\rm H_2}}
\def\rhohi{\rho_{\rm HI}}
\def\nhtwo{n_{\rm H_2}}
\def\nhi{n_{\rm HI}}

We then calculated the distribution of molecular fraction (Elmegreen 1993)
corresponding to figure 1.
The molecular fraction is defined by
\begin{equation}
\fmol
= \frac{\rhohtwo}{\rhohi + \rhohtwo}
= \frac{2  \nhtwo}{ \nhi + \nhtwo},
\end{equation}
where $\rhohtwo$, $\rhohi$, $\nhtwo$ and $\nhi$ are the mass and number
densities of molecular and \HI\ gases, respectively.
We used a method described by Sofue et al. (1995) and Honma et al. (1995) to
calculate the galaxy-scale molecular fraction; they investigated
molecular fronts in spiral galaxies using a phase-transition model
proposed by Elmegreen (1993).
In this model, the molecular fraction is determined by
three parameters; the interstellar pressure, $P$, the UV radiation field,
 $U$, and the metallicity, $Z$.
For $U$ and $Z$, we adopted an exponential function of
galactocentric radius, and calculated $\fmol$ for corresponding gas
pressure, $P$, which is determined by the gas density in each cell.

Figure 3 shows the result of a numerical simulation of the molecular fraction.
The inner few kpc region is dominated by molecular gas, where
the molecular fraction is as large as $\sim 70-90$\%.
Also, the molecular fraction increases suddenly at the galactic shocks,
which are already deformed from symmetric arms.
Thus, the simulation has revealed that highly deformed inner
molecular arms can be produced by ram-pressure disturbances on the
inter-arm low-density regions.

--- Fig. 3 ---

\subsection{Comparison with Observations and Wind Velocities}
We now compare the results with \HI\ and CO-line observations of the
Virgo galaxies, NGC 4254 and NGC 4654.
The heliocentric radial velocity of NGC 4254, 2407 \kms,
is about 1100 \kms\ different from that of NGC 4486 (M 87),
the center of the Virgo cluster, 1282 \kms\
(de Vaucouleurs et al. 1991).
Also, by comparing with the lopsided \HI\ distribution of
NGC 4254 (Phookun et al. 1993) and the ram pressure simulation on
galaxies (Abadi et al. 1999), we can exclude the possibility of
any face-on motion of NGC 4254. Assuming that NGC 4254's orbit
is inclined by $45^{\circ}$ from the line of sight, we may estimate
the velocity of motion of NGC 4254 in the Virgo cluster as being 
$\sim$ 1500 \kms.
Then, the rotation velocity of NGC 4254 ($\sim$ 150 \kms) is negligible
compared to the wind velocity, so that the \HI\ tail grows toward downstream.
The location of a prominent spiral arm (Iye et al. 1982) and the
direction of rotation are consistent with our numerical simulation.
However, if we assume a slower wind velocity, e.g., on the order of, or smaller
than, $\sim$ 750 \kms, the simulation cannot reproduce the observed features.

Our simulation for moderate ICM velocity, which predicts an off-center bar of
dense gas tilted toward the ICM wind, is consistent with the observations of
NGC 4654 in \HI\ (Phookun, Mundy 1995) and CO (Sofue et al. in preparation).
Our simulations show that spiral arm in the upstream side becomes
stronger than that in the downstream side.
Considering the location of the prominent optical arm (Frei et al. 1996)  and
an elongation of the observed molecular bar and \HI\ tail, we find that the 
direction of motion of NGC 4654 is toward the northwest, 
with a significantly high
velocity compared to the velocity dispersion of the Virgo cluster.
Taking it into account that the rotation velocity of NGC 4654 is relatively
slow, the velocity of NGC 4654 in the Virgo cluster would be greater
than  1000 \kms, although the heliocentric radial velocity,
1054 \kms, is close to Virgo's central velocity.

\section{Discussion}

We have considered the ram-pressure effects of a 
mild ICM wind on gaseous disks in cluster galaxies.
Galaxies in the central region of the Virgo galaxies show \HI\ deficiency,
where the molecular gas fraction is higher than that of 
the galaxies in the outer region of the cluster (Kenney et al. 1990).
This fact has been naturally explained by the ram-pressure stripping
of the \HI\ gas: the ram effects are negligible on the inner
molecular disks, while it is crucial for the \HI\ outer disk and envelops.
On the other hand, it is also known that ram-affected Virgo galaxies show
asymmetric CO gas distributions (Kenney et al. 1990; Sofue et al. in  
preparation).

In the present paper, we have shown that the orbits of the inter-arm ISM
are disturbed by the ram pressure of the ICM wind, even if the wind is 
mild and not strong enough to strip the gas disk.
The disturbed inter-arm ISM causes highly asymmetric molecular arms in
the inner  few kpc of the disk.
The 3D simulation by Vollmer et al. (2000, 2001) for a similar wind condition
shows that the inner disk within 10 kpc radius is not stripped, but suffers 
from perpendicular disturbances to the disk plane.
Vertical displacements may result in off-plane molecular structure in the
inner few kpc disk, in addition to the asymmetric arms as simulated here
using the 2D scheme.
Also, such 3D effects as the Kelvin--Helmholtz instability by the shearing 
motion between the disk and ICM (e.g. Mori and Burkert 2000) cannot be 
touched upon by the present 2D simulation.
Therefore, detailed 3D simulations  will be crucial to thoroughly understand
the inner ram effect in more detail, while the present 2D results can tell 
us about some essential mechanism to cause the non-axisymmetric molecular 
structures in the inner disk.

However, in the cases of much stronger winds, the 2D assumption cannot be 
applied in any way, and 3D treatment of stripping is crucial. 
In fact, ram effect by a wind with a
pressure much higher than that considered in this paper has been simulated 
by Quilis et al. (2000) with a 3D hydrodynamical simulation; they have shown 
that the ISM in a galaxy is completely stripped within one galactic rotation.

\vskip 10mm
\noindent{\bf References}
\vskip 5mm

\def\ref{\hangindent=1pc \noindent}

\ref Abadi, M. G., Moore, B.,
	\& Bower, R. G., 1999, MNRAS, 308, 947

\ref Balsara, D, Livio, M, \& O'Dea, C P. 1994, ApJ, 437 83

\ref Blondin, J. M., \& Lufkin, E. A., 1993, ApJS, 88, 589

\ref Cayatte, V., van Gorkom, J. H., Balkowski, C., \& Kotanyi, C. 
1990, AJ, 100, 604  

\ref Colella, P., \& Woodward, P. R.,
	1984, Journal of Comp. Phys., 54, 174

\ref de Vaucouleurs, G.,
	de Vaucouleurs, A., Corwin, H. G. Jr., Buta, R. J., Paturel, G.,
	\& Fouqu\'e P. 1991, Third Reference Catalog of
	Bright Galaxies (New York: Springer-Verlag)

\ref Elmegreen, B. G. 1993, ApJ, 411, 170

\ref Farouki, R. \& Shapiro, S. L. 1980, ApJ, 241, 928

\ref Frei, Z., Guhathakurta, P.,
	Gunn, J. E., \& Tyson, J. A., 1996, AJ, 111, 174

\ref Gaetz, T. J.,  Salpeter, E. E.,  Shaviv, G. 1987 ApJ 316 530.

\ref Honma, M., Sofue, Y., \& Arimoto, N. 1995 A\&A, 304, 1

\ref Iye, M., 	Okamura, S., Hamabe, M., \& Watanabe, M. 1982, ApJ, 256, 103

\ref Kenney, J. D. \& Young, J. S. 1988, ApJS., 66, 261

\ref Kenney, J. D. P., Young, J. S., Hasegawa, T., \& Nakai, N. 
1990, ApJ, 353, 460

\ref Kenney, J. D. P., \& Koopmann, R. A. 1999, AJ, 117, 181

\ref Kritsuk, A. G., 1984, Astrophysics, 19, 263 

\ref Mori, M.,  \& Burkert, A. 2000, ApJ,  538, 559

\ref Phookun, B., Vogel, S. N., \& Mundy, L. G. 1993, ApJ, 418, 113

\ref Phookun, B., \& Mundy, L. G., 1995, ApJ, 453, 154

\ref Quilis, V.,  Moore, B.,  \& Bower, R. 2000, Science, 288, 1617

\ref Sofue, Y. 1994, ApJ, 423, 207

\ref Sofue, Y., Honma, M., \& Arimoto, N., 1995, A\&A, 296, 33.

\ref Toomre, A. 1981, in The Structure and Evolution of Normal 
	Galaxies, ed S. M. Fall \& D. Lynden-Bell
	(Cambridge: Cambridge Univ. Press), 111

\ref Tosa, M. 1994, ApJ,  426, L81

\ref Vollmer, B.,  Marcelin, M., Amram, P., Balkowski, C., 
	Cayatte, V., \& Garrido, O.   2000, A\&A, 364, 532

\ref Vollmer, B., Cayatte, V., Balkowski, C., \& Duschl, W.J. 2001, ApJ,
561, 708

\vskip 20mm

Figure Captions

\vskip 5mm

Fig. 1. Schematic illustration of the ram-deformation mechanism to cause
asymmetric inner molecular arms. 
Orbits (thin dashed arrows) of low-density inter-arm gas (A and A$'$) are 
disturbed by the ram pressure of the ICM wind (thin arrows).
The gas on the distorted orbits (thick arrows) encounters density 
waves (dashed spirals) at different places (B and B$'$) from those expected 
for undisturbed orbits (dashed spirals), and produces deformed dense 
gaseous arms (thick spirals). 

\vskip 5mm
Fig. 2. Snapshots of the density distribution in the ram-pressure models
after $1.3$--$1.7\times 10^8$ yr.
The labels indicated the adopted parameters.
Alphabets: ICM velocity, $\vec{V}_{\rm{ICM}} = $ (a) 530, (b) 1000,
and (c) 1500 \kms.
Roman numbers: ICM density $n_{\rm{ICM}} = $ (i) $1 \times 10^{-4}$,
(ii) $ 5 \times 10^{-4}$,  and (iii) $ 1 \times 10^{-3}$ cm${}^{-3}$.
The color-density key is shown at the bottom, which is common to all
snapshots.
The rotation direction of gases and a bar potential is counterclockwise,
and ICM wind brows from left to right. The strength of the bar
$\varepsilon$ is 0.10 in all models.
Weak straight waves toward upper and right boundaries in b-(ii, iii) 
and c-(ii, iii) are artifact due to numerical reflection at the boundaries

\vskip 5mm
Fig. 3.  Same as figure 2, but showing the distribution of
molecular fraction, $\fmol$.

\end{document}